# Threshold Switching in Vertically Aligned MoS$_2$/SiO$_x$ Heterostructures based on Silver Ion Migration

*Jimin Lee [1], Rana Walied Ahmad [2], Sofía Cruces [1], Dennis Braun [1], Lukas Völkel [1], Ke Ran [3,4,5], Joachim Mayer [4,5], Stephan Menzel [2], Alwin Daus [1,6,\*], and Max C. Lemme [1,3,\*]*

[1] Chair of Electronic Devices, RWTH Aachen University, Otto-Blumenthal-Str. 25, 52074 Aachen, Germany.

[2] JARA-Fit and Peter Grünberg Institute (PGI-7), Forschungszentrum Jülich GmbH, Wilhelm-Johnen-Str., 52428 Jülich, Germany.

[3] AMO GmbH, Advanced Microelectronic Center Aachen, Otto-Blumenthal-Str. 25, 52074 Aachen, Germany.

[4] Central Facility for Electron Microscopy, RWTH Aachen University, Ahornstr. 55, 52074 Aachen, Germany.

[5] Ernst Ruska-Centre for Microscopy and Spectroscopy with Electrons, Forschungszentrum Jülich GmbH, Wilhelm-Johnen-Str., 52428 Jülich, Germany.

[6] Institute of Semiconductor Engineering, University of Stuttgart, Pfaffenwaldring 47, 70569 Stuttgart, Germany.

**ABSTRACT**

Threshold switching (TS) is a phenomenon where non-permanent changes in electrical resistance of a two-terminal device can be controlled by modulating the voltage bias. TS based on silver (Ag) conductive filaments has been observed in many materials, including layered two-dimensional (2D) transition metal dichalcogenides (TMDs). 2D TMDs are particularly promising for metal ion movement due to their van der Waals (vdW) gaps between their sheets, facilitating ion migration and filament formation without disturbing covalent chemical bonds. In this work, we demonstrate the heterostructure growth of vertically aligned molybdenum disulfide (VAMoS$_2$) with an amorphous silicon oxide (SiO$_x$) layer on top after sulfurization. We show that Ag ions migrate through this material stack, enabling TS. Our Ag/SiO$_x$/VAMoS$_2$/gold (Au) devices exhibit TS at low voltages of ~0.63 V, with high on-state currents over 200 µA and stable switching exceeding ≥$10^4$ cycles. Moreover, we identify two rate-limiting steps for filament formation through a physics-based dynamical model and simulate the switching kinetics. Our devices show a fast on-switching time of 311 ns and spontaneous relaxation in 233 ns. These findings deepen the understanding of $SiO_x$/$MoS_2$-based RS devices and demonstrate the promise for applications in emerging memories and neuromorphic computing systems.

# 1. Introduction

Threshold switching (TS), also known as volatile resistive switching (RS), is a phenomenon in which electrical resistance temporarily changes in response to an applied electrical field[1]. TS devices have become highly relevant for emerging technologies, e.g., to be used as memory selectors, as artificial neurons and synapses in neuromorphic computing, and as true random number generators for security applications[2–9]. Memristive devices are typically composed of a metal-insulator-metal structure. They show TS primarily due to filament formation[10–12] achieved by ionic transport[13–15].

Recently, TS driven by the formation and disruption of conductive filaments from highly diffusive metal electrodes, such as silver (Ag) or copper (Cu) in two-dimensional (2D) materials, has gained considerable attention[16–18]. Reliable volatile RS, including high-speed switching and high endurance, is enabled by thin or weak filaments of the active metal ions in the switching layer[19,20]. The relaxation process, where filaments dissolve due to surface energy minimization or thermal diffusion, is influenced by the properties of the filament and the switching layer[20–23]. Material engineering strategies, such as a defected graphene layer at the Ag/insulator interface, have also been reported to influence the filament formation process[3,18]. In addition, the active metals enable the switching dynamics to be modulated by voltage amplitude, pulse duration, and temperature[5,8,18].

Ag or Cu filamentary threshold switches combined with 2D transition metal dichalcogenides (TMDs) are investigated for advanced brain-inspired neuromorphic hardware with fast switching speeds, high integration density, and low power consumption[10,24–27]. Among the 2D TMDs, molybdenum disulfide ($MoS_2$) is one of the most extensively studied switching layers in memristive devices[28]. $MoS_2$ films can be synthesized on wafer scale by thermally assisted conversion (TAC) from molybdenum (Mo) thin films into lateral or vertical $MoS_2$ layers via sulfurization[29–31]. Vertically oriented $MoS_2$ layers are particularly promising for crossbar architectures, which rely on vertical electrical connections [11,12,15,32]. Their vertically oriented van der Waals (vdW) gaps facilitate ion movement between two sandwiched electrodes for RS behavior[11,15]. In fact, $MoS_2$-based volatile RS devices with an Ag active electrode have shown low threshold voltages from 0.18 V to 0.35 V and a high ON-OFF ratio of $10^6$[12,16,33]. However, their time-dependent behavior has rarely been studied, in contrast to their widely investigated oxide-based TS devices[3,5,8]. Increased understanding of the switching dynamics in $MoS_2$-based RS devices is key for their potential application.

In this work, we present TS devices that consist of an amorphous, substoichiometric silicon oxide ($SiO_x$) layer on vertically aligned $MoS_2$ ($VAMoS_2$) below an active Ag electrode. The $SiO_x$ layer was formed on top of $VAMoS_2$ during the sulfurization process. We show that Ag ions can travel through the material stack by applying long and short voltage pulses, leading to volatile RS. The switching performance is improved at lower switching voltages with a thinner $SiO_x$ layer on top of the $VAMoS_2$. The switching dynamics are analyzed theoretically using the JART ECM simulation model[34,35], which considers physics-based rate-limiting ionic processes.

## 2. Results and Discussion

2.1 Device fabrication and material characterization of $SiO_x$/$MoS_2$-based RS devices

The two-terminal vertical devices studied in this work comprise a stack of $SiO_x$ and $VAMoS_2$ layers sandwiched between an active Ag top electrode (TE) and an inert gold (Au) bottom electrode (BE). Figures 1a and b show a schematic of the TS device and an optical microscopy image, respectively. Figure 1c shows the fabrication process beginning with the preparation of inert Au BEs on Si substrates with 300 nm thermally grown silicon dioxide ($SiO_2$) using photolithography, electron-beam evaporation, and lift-off techniques. 6 nm Mo thin films were deposited on the pre-patterned Au BEs and subsequently converted into $VAMoS_2$ layers by TAC in a sulfur atmosphere in a chemical vapor deposition (CVD) system[15]. We fabricated $SiO_x$/$VAMoS_2$ devices with two different $SiO_x$ layer thicknesses by placing the substrate either upright against the boat wall for thicker $SiO_x$ and face-down on two overlapping boats for thinner $SiO_x$ inside the CVD quartz tube (Figure 1d). This resulted in 60 nm and 10 nm $SiO_x$ on top of the $VAMoS_2$, as confirmed by ellipsometry. Finally, the active Ag TEs were deposited and structured similarly to the BEs (see Experimental Section and Supporting Information (SI) Figure S1 for further details).

The device compositions and structures were verified using cross-sectional transmission electron microscopy (TEM), which confirmed the presence of an amorphous $SiO_x$ layer on $VAMoS_2$ between the Ag and Au electrodes (Figures 2a, b). The vertical orientation of the nanocrystalline $MoS_2$ layers was further verified by high-resolution TEM (HRTEM, Figure 2c). Raman analysis was conducted on the as-grown $MoS_2$ film on the Au BE surface for both samples (SI Figure S2). Raman spectra of the sulfurized $MoS_2$ films show the characteristic $E^1_{2g}$ and $A_{1g}$ peaks at 382 cm$^{-1}$ and 407 cm$^{-1}$ in both devices. The Raman peaks correspond to the in-plane ($E^1_{2g}$) and out-of-plane ($A_{1g}$) vibrational modes, respectively[36,37], and confirm the synthesis of the 2H-MoS$_2$ phase. The data also provide insight into the vdW layer orientation, as their peak intensity ratio ($E^1_{2g}/A_{1g}$) of approximately 0.4, observed in both cases, indicates the formation of vertical MoS$_2$ layers, as reported in the literature[38]. The vertical MoS$_2$ growth from 6 nm Mo thin films is consistent with the literature, where vertical vdW orientation was reported for initial Mo thicknesses exceeding 3 nm (see Figure 2d for details)[12,30–32,39–43].

The presence of the amorphous SiO$_x$ layer was confirmed by high-angle annular dark-field (HAADF) imaging (Figure 2e) and energy-dispersive X-ray spectroscopy (EDXS) elemental mapping (Figure 2f) of the 10 nm SiO$_x$/VAMoS$_2$ device (see SI Figure S3 for the 60 nm SiO$_x$/VAMoS$_2$ device). We attribute the SiO$_x$ layer formation to atomic diffusion of Si and O from the SiO$_2$/Si substrate at high temperatures during the Mo sulfurization process[32,44], potentially enhanced by the vdW gaps in the VAMoS$_2$ layer.

2.2 DC electrical performance

The SiO$_x$/VAMoS$_2$ devices initially exhibited a high resistance of more than 25 GΩ, which can be explained by the presence of the insulating SiO$_x$ in the material stack. The RS characteristics of thick (60 nm) and thin (10 nm) SiO$_x$/VAMoS$_2$ devices were measured by 30 consecutive direct-current (DC) current–voltage (*I–V*) sweeps with a current compliance of $I_{cc}$ = 1 μA (Figures 3a, b). Both devices showed abrupt switching from the high resistance state (HRS) to the low resistance state (LRS) at certain positive on-threshold voltages ($V_{t,on}$) applied to the Ag top electrode. The LRS was maintained for a positive bias beyond $V_{t,on}$. The transition from LRS back to HRS occurred during the reverse voltage sweep below the hold voltage ($V_{hold}$). Repeatable volatile RS was observed with an average $V_{t,on}$ of ~1.66 V for the 60 nm SiO$_x$/VAMoS$_2$ device and ~0.63 V for the 10 nm SiO$_x$/VAMoS$_2$ device, demonstrating that a thinner SiO$_x$ layer on VAMoS$_2$ requires a lower switching voltage.

TEM imaging revealed traces of Ag in the SiO$_x$ layer in the HRS (~1 MΩ) after 42 switching cycles (Figure 3c). This suggests that applying a positive voltage to the TE induces Ag diffusion through the switching layer and the formation of a conductive filament bridging the top and bottom electrodes[16]. However, the Ag ions visible in the TEM do not form a complete filament. Instead, the conductive filament required for the LRS has partly dissolved and ruptured as the voltage bias was removed due to interfacial energy minimization [12,19]. The volatile switching in our devices is thus governed by the formation and breakage of Ag conductive filaments within the SiO$_x$/VAMoS$_2$ stack, enabled by Ag ion movement in the SiO$_x$ layer and along the vdW gaps of vertically oriented MoS$_2$ layers. Similar non-volatile switching with ion movement along the grain boundaries has been observed in laterally oriented MoS$_2$ layers[10,17,26,31].

We evaluated the variability of the devices based on SiO$_x$ thickness by statistically analyzing $V_{t,on}$ and $V_{hold}$ data from 30 consecutive cycles (Figures 3d, e). The 60 nm SiO$_x$/VAMoS$_2$ device exhibited larger cycle-to-cycle standard deviations of 0.39 V for $V_{t,on}$ and 0.21 V for $V_{hold}$, compared to the 10 nm SiO$_x$/VAMoS$_2$ device with standard deviations of 0.16 V for $V_{t,on}$ and 0.09 V for $V_{hold}$, respectively. The thinner SiO$_x$ layer on VAMoS$_2$ reduces variability and enables lower switching voltages. Additional DC I–V sweeps were performed on multiple devices to quantify the device-to-device variation. The results are shown in Figure 3f, which plots $V_{t,on}$ and $V_{hold}$ of the initial I–V traces of ten 60 nm and twenty 10 nm SiO$_x$/VAMoS$_2$ devices. The individual I–V sweeps corresponding to these data are provided in SI Figures S4 and S5. The 60 nm SiO$_x$/VAMoS$_2$ devices exhibited standard deviations of 0.53 V for $V_{t,on}$ and 0.34 V for $V_{hold}$, while the 10 nm SiO$_x$/VAMoS$_2$ devices showed lower deviations of 0.21 V for $V_{t,on}$ and 0.13 V for $V_{hold}$, respectively. The lower cycle-to-cycle and device-to-device variability in the latter, combined with the lower switching voltages, favors thin SiO$_x$ layers on VAMoS$_2$ for applications such as artificial neuron circuits[18]. The cycle-to-cycle variation for the 10 nm SiO$_x$/VAMoS$_2$ device was further evaluated using cumulative probability analysis over 384 consecutive cycles (Figure 3g), with all *I-V* sweeps for the device plotted in SI Figure S6.

2.3 Pulsed voltage characteristics

We conducted pulsed voltage measurements to further elucidate the switching dynamics of the 10 nm SiO$_x$/VAMoS$_2$ devices, selected for their favorable DC switching behavior. We analyzed the transient current response to 100 µs voltage pulses with amplitudes ranging from $V_{pulse}$ = 1.5 V to 4.5 V. The devices switched to the LRS after a certain time which depended on $V_{pulse}$. After the pulse, the device returned to the HRS, and the state was read out with a read pulse of 0.3 V for 3 µs. Two typical transient current responses are shown in Figures 4a, b (all data are available in SI Figure S7). The on-switching time ($t_{on}$) was determined by extracting the midpoint value of the first significant non-capacitive current rise in the double logarithmic current-time (log(*I*)-log(*t*)) curves. This midpoint value also represents the steepest slope during this specific transient current rise. Figure 4c shows the extracted $t_{on}$ values as a function of $V_{pulse}$, with an inset illustrating the voltage pulse waveform. As $V_{pulse}$ increases, the log($t_{on}$) decrease is mostly linear. However, at higher $V_{pulse}$ values, the rate of decrease in log($t_{on}$) reduces, resulting in a progressively flatter log($t_{on}$)-$V_{pulse}$ curve, which indicates reduced sensitivity of the switching time change to further increased voltages. Similar observations were reported in volatile and non-volatile electrochemical metallization devices, where the switching kinetics are divided into three regimes based on the respective ionic rate-limiting step: (I) nucleation limited, (II) electron transfer limited, and (III) electron transfer and ion migration limited[8,34].

We applied a well-established simulation model developed by Menzel *et al.* and Ahmad *et al.*[34,35] to gain a deeper understanding of the physical processes limiting the switching time in our Ag/SiO$_x$/VAMoS$_2$/Au devices. The full model is explained in SI Section S7. The simulation results are plotted as a solid line with the experimental data in Figure 4c. The fitting parameters are detailed in SI Table S1. Two distinct regimes are evident from the data: At voltages below 3 V, the sharp linear slope on the logarithmic scale corresponds to the electron transfer limited regime. Above 3 V, the flatter slope indicates that the rate-limiting factor shifts to a mixed regime of electron transfer and silver ion migration. Therefore, the voltage that drops across the device determines which of the ionic processes is/are the slowest and, thus, rate-limiting. This insight is obtained by analyzing the physics-based model, which describes the transient evolution of the gap between the active electrode and the filament tip and the transient evolution of voltage drops across the various ionic processes [34,35]. The model also includes a nucleation limited regime, which can typically be accessed with rather low voltages applied for rather long pulse durations, but which we did not see in this work's experimental data.

Next, we explored short voltage pulses of 1 µs to switch the SiO$_x$/VAMoS$_2$ devices. The voltage-dependent transients in Figure 4d show a significant increase in on-state current upon 1 µs pulses of different voltages, up to 200 µA for 6 V. Such short pulses lead to reduced stress, which typically enhances the device endurance, a key performance parameter for TS devices besides fast switching time and high on-state current. We tested the endurance by applying short 2 µs voltage pulses of 4.5 V to switch a device from the HRS to the LRS, then returning it to the HRS, followed by a pulse of 0.1 V for 3 µs to read out the off-state current. The 10 nm SiO$_x$/VAMoS$_2$ device exhibits stable switching behavior after 10$^4$ cycles with an average ON-OFF ratio of $10^3$, as shown in Figure 4e. Additional DC resistance measurements were taken after every 10 consecutive voltage pulses to further validate the reliability of the device's volatility. The DC resistance values are plotted together with the extracted HRS and LRS resistance values in SI Figure S8, confirming consistent returns to the HRS after each set of 10 pulses. A pulsed endurance test was also performed without the DC resistance measurements to eliminate any potential influence from them. This device also demonstrated continued robust endurance of ~5800 cycles, with average resistance values of 166 kΩ for the LRS and 5 MΩ for the HRS (see SI Figure S9).

Furthermore, we measured the $t_{on}$ and the relaxation time ($t_r$) of a 10 nm $SiO_x$/VAMoS$_2$ devices as a function of $V_{pulse}$ for a fixed pulse time of 2 µs over 30 consecutive pulses. $V_{pulse}$ values between 3.5 V and 5.0 V were chosen, where the switching time is determined by the electron-transfer and ion migration, enabling a rapid transition to the LRS. One example of a measured time-resolved current response to a 5 V pulse is shown in Figure 4f. Here, $t_{on}$ is defined as the time required to reach 90% of the LRS ON-current ($I_{on}$). $t_r$ is defined as the time needed to decay from $I_{on}$ to 10% of the difference between $I_{on}$ and the HRS OFF-current ($I_{off}$)[45]. All extracted $t_{on}$ and $t_r$ are plotted in Figures 4g and h, with the inset in h showing the voltage pulse waveform. $t_{on}$ decreases slightly with increasing $V_{pulse}$, similar to the behavior observed with the longer voltage pulses of 100 µs. The device then returns to the HRS through a self-relaxation process while the read voltage is applied. Unlike $t_{on}$, $t_r$ increases with higher $V_{pulse}$. We assume that the stronger electric fields cause the formation of thicker, more robust filaments, which take longer to rupture. This assumption is supported by recent studies on oxide memristive devices[5,23]. The relaxation process thus reflects the history of the device, with $t_r$ revealing whether the formed filament was thick or thin[8].

Our 10 nm SiO$_x$/VAMoS$_2$ device exhibits a fast switching of $t_{on}$ = 311 ns and a spontaneous relaxation time of $t_r$ = 233 ns using a voltage pulse of 5 V/2 µs and a read pulse of 0.3 V/3 µs. $t_{on}$ and $t_r$ are critical parameters for operating memristive devices and can be adjusted by the voltage pulse parameters in our devices. Their tunability can be used for versatile applications in electronics, ranging from selectors for memories to synaptic devices for neuromorphic computing[3–6,8]. Our device shows promise for selector applications, which require high on-state currents and short relaxation times[4,46]. However, there is a tradeoff, as increasing the on-state current tends to increase the relaxation time.

Table 1 compares the TS performance of the present SiO$_x$/VAMoS$_2$ devices with data from previously reported studies in the literature[3,4,12,33,47]. Our devices demonstrate significantly lower switching voltages and faster switching speeds, particularly faster relaxation times, compared to pure SiO$_x$ or SiO$_2$-based devices. Our SiO$_x$/VAMoS$_2$ devices exhibit 10 times higher $V_{hold}$ than the SiO$_2$ devices[47] although they have similar $V_{t,on}$, which implies that weaker Ag filaments are formed. We attribute this to the rapid formation and dissolution of conductive filaments within the switching stack, particularly in the VAMoS$_2$ layer, where the vdW nature of MoS$_2$ enables ionic transport but also limits the filament growth. The relaxation time is mainly governed by the material stack, due to spontaneous conductive path rupture driven by atomic and ionic mass transport[21]. The SiO$_x$ layer is known to promote the formation of strong conductive Ag filaments[48,49]. However, the presence of the VAMoS$_2$ layer in direct contact with the Au BE appears to suppress this strong filament growth in our SiO$_x$/VAMoS$_2$ devices. When comparing the SiO$_x$/VAMoS$_2$ and SiO$_x$-only devices, we found that the incorporation of the VAMoS$_2$ layer leads to more stable switching behavior by the confined formation of filaments. A detailed comparison of their pulsed voltage responses is presented in SI Figure S10. Our findings highlight the potential to achieve fast switching by integrating a SiO$_x$ layer with 2D TMDs and provide insight into switching dynamics for volatile RS devices.

## 3. Conclusion

We investigated Ag-based TS using a material stack comprising thin SiO$_x$ layers on VAMoS$_2$ in a Ag/SiO$_x$/VAMoS$_2$/Au configuration. Our devices demonstrate highly repeatable volatile RS at low switching voltages of 0.63 V, with a high on-state current exceeding 200 μA for $V_{pulse}$ = 6 V and robust pulsed endurance over 10$^4$ cycles. Furthermore, our devices show fast switching filament formation and dissolution speeds with both switching and relaxation times in a few hundred nanoseconds for $V_{pulse}$= 5 V. We analyzed the switching kinetics of the devices using the JART ECM model and identified the experimentally observed physical processes of electron-transfer limited and electron-transfer/ion migration limited switching. These findings advance the understanding of SiO$_x$/VAMoS$_2$-based RS devices and help pave the way for their practical use in emerging memories and neuromorphic computing systems.

## 4. Experimental Section

*Device Fabrication and Material Growth*: The Ag/SiO$_x$/VAMoS$_2$/Au cross-point devices, with dimensions of 4 µm x 4 µm, were fabricated on 300 nm SiO$_2$/Si substrates. First, the substrates were cleaned using acetone, isopropanol (IPA) and O$_2$ plasma to remove surface contaminants. Bottom electrodes (BEs) were defined by photolithography (using a Mask aligner EVG420 from EV GROUP) and deposited with a 5 nm Ti adhesion layer and a 50 nm Au layer by electron-beam evaporation (Pfeiffer Balzers PLS 500), followed by lift-off in dimethyl sulfoxide (DMSO) heated to 80 °C. Next, a 6 nm thick patterned molybdenum (Mo) film was formed on the BEs by photolithography, direct-current sputtering (CREAMET S2, CREAVAC GmbH), and lift-off. A Mo target with a purity of 99.95% from EVOCHEM was used for the sputter process. The sputtering was conducted with a power of 100 W under an Ar flow and a chamber pressure of ~3 × 10$^{-3}$ mbar. The sulfurization process was performed in a chemical vapor deposition (CVD) tube furnace (RS 80/300, Nabertherm GmbH), where the patterned Mo film on Au/SiO$_2$/Si substrate was heated in a sulfur atmosphere. This process was conducted at 800 °C for 30 minutes with an Ar flow rate of 20 sccm as the carrier gas at a pressure of 9.6 x 10$^{-2}$ mbar to achieve a vertical orientation of the MoS$_2$ layers. Sulfur powder was placed upstream in the CVD furnace and heated to 150 °C. Sulfur diffused into the Mo film, creating ~20 nm thick MoS$_2$ layers. Finally, 30 nm Ag top electrodes (TEs) with a 30 nm Au capping layer were deposited using photolithography, electron-beam evaporation, and lift-off processes. The complete fabrication process flow and the sulfurization process are illustrated in Figures 1c, d. Top-view optical microscopy images of the thick and thin SiO$_x$/VAMoS$_2$ cross-point devices are shown in SI Figures S1a, b, respectively.

*Material Characterization*: Transmission electron microscopy (TEM) was employed to investigate the cross-sectional device structure, layer orientation, and chemical composition. TEM lamellas were prepared by focused ion beam (FIB) using an FEI Strata400 system. TEM imaging and energy-dispersive X-ray spectroscopy (EDXS) were conducted by JEOL JEM F200 at 200 kV. Raman spectroscopy was performed on the as-grown $MoS_2$ layer to analyze the phase formation. The Raman measurement was taken using a WITec alpha 300R Raman system in mapping mode with a 532 nm excitation laser at 1 mW power and a 1800 g/mm grating.

*Electrical Characterization*: Switching properties of the devices were measured at room temperature using a Lakeshore probe station connected to a semiconductor parameter analyzer (Keithley 4200S-SCS from a Tektronix company), equipped with two source measure unit cards (Keithley 4200-SMU) and preamplifiers (Keithley 4200-PA). A voltage stimulus was applied to the Ag top electrode while the Au bottom electrode was grounded. Under an applied electric field, Ag ions migrate through the material stack, namely the $SiO_x$ layer and the $VAMoS_2$ layers, modulating the conductivity. The applied voltage was monitored in parallel on channel 2 from the Ag top electrode, and the output current was measured over time on channel 1 from the Au bottom electrode. For DC Current–Voltage (*I–V*) sweep measurements, the voltage was swept from 0 V to a positive maximum 3 V and back to 0 V, with the current limited by an external current limiter within the Keithley 4200-SCS. Pulsed voltage measurements were conducted using the pulse measure units (Keithley 4225-PMUs).

**Acknowledgements**

This research was supported in part by the German Federal Ministry of Education and Research (BMBF) through NEUROTEC2 (16ME0398K, 16ME0399 and 16ME0400) and NeuroSys as part of the initiative "Clusters4Future" (03ZU1106AA, 03ZU1106AB) projects, by the European Union's Horizon Europe research and innovation program via CHIPS-JU under the ENERGIZE project (101194458), and by the Deutsche Forschungsgemeinschaft (DFG) within the project MEMMEA in SPP 2262 MemrisTec (441918103) and the Emmy Noether Programme (506140715).

**Conflict of Interest**

The authors declare no conflict of interest.

**Data Availability Statement**

The data that support the findings of this study are available from the corresponding author upon reasonable request.

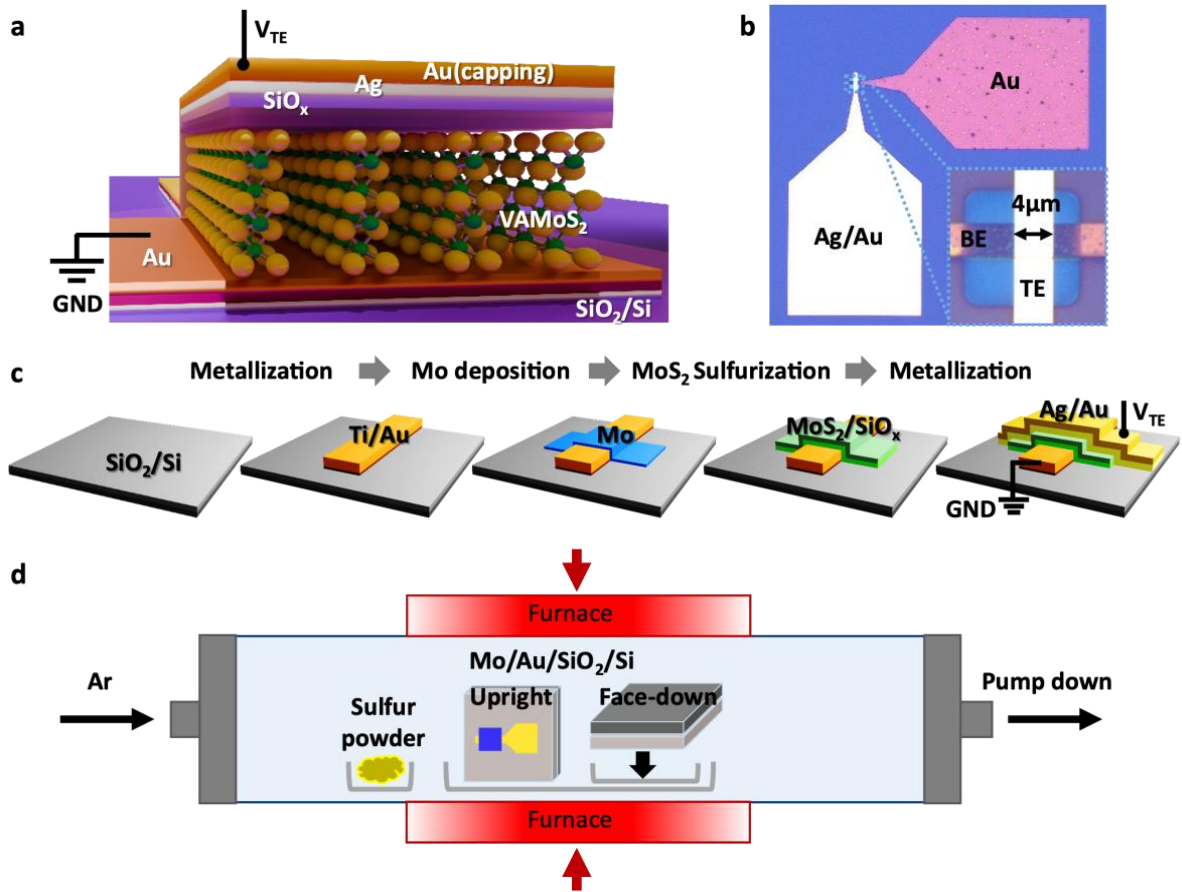

**Figure 1. Ag/SiO$_x$/VAMoS$_2$/Au device structure and fabrication process.**

(a) Schematic drawing of the TS device. (b) Top-view optical microscopy image of the fabricated device, with the inset showing a zoomed image of the 16 μm$^2$ area where the top (TE) and bottom (BE) electrodes intersect. (c) Schematic of the fabrication process flow for a RS cross-point device. (d) Schematic of the sulfurization process for VAMoS$_2$ film in a horizontal tube furnace, with sulfur powder positioned upstream and sputtered Mo thin film on pre-patterned Au on SiO$_2$/Si substrate at the center of the tube.

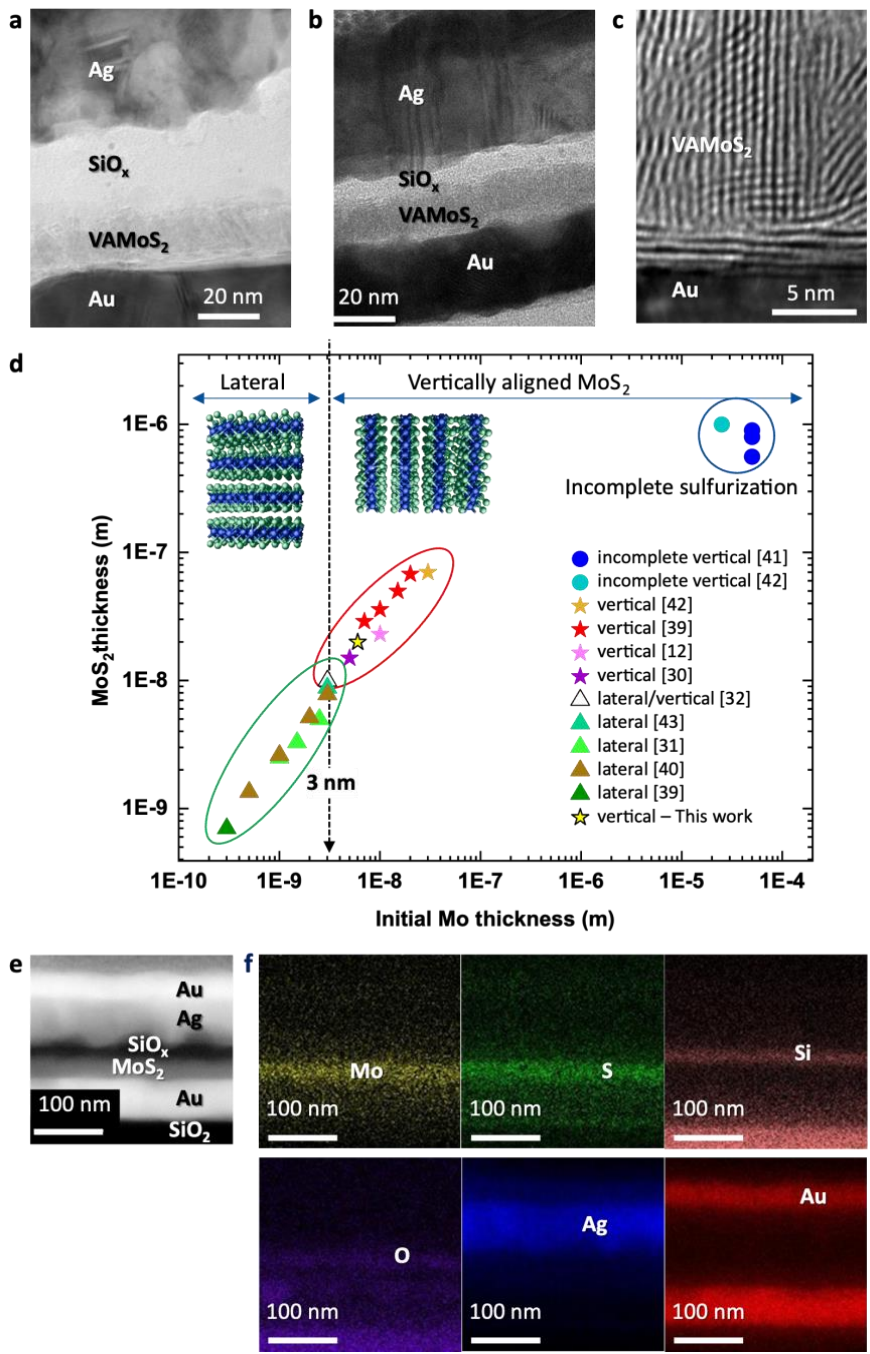

**Figure 2. Cross-sectional analysis of the device structure and material characterization.**

(a,b) Cross-sectional TEM images displaying the vertical stack of $SiO_x$/VAMoS$_2$ between the top and bottom electrodes, showing devices with 60 nm (a) and 10 nm (b) $SiO_2$ layers, respectively. (c) An enlarged HRTEM image of VAMoS$_2$ layers grown on Au. (d) MoS$_2$ orientation as a function of initial Mo layer thickness. When the initial Mo thickness is below 3 nm, laterally aligned MoS$_2$ layers are formed. As the Mo thickness increases, vertically aligned MoS$_2$ layers begin to dominate. In the bulk, incomplete sulfurization occurs after a certain amount of vertical layer formation[12,30–32,39–43]. (e,f) Cross-sectional HAADF image of the 10 nm SiO$_x$/VAMoS$_2$ device (e), and the corresponding EDXS elemental mapping of the device stack (f), showing Mo, S, Si, O, Ag, and Au.

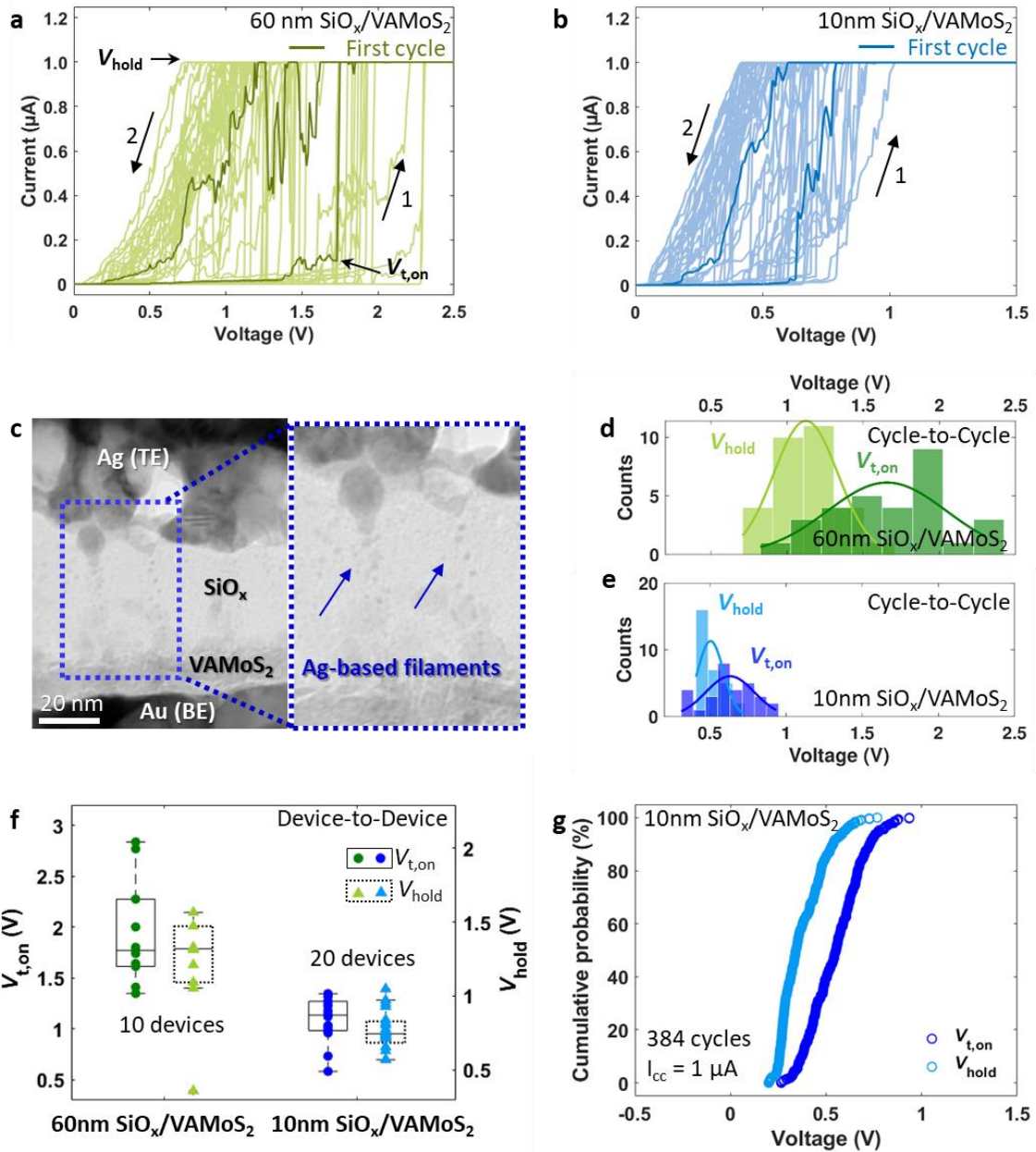

**Figure 3. DC current-voltage (*I-V*) characteristics of Ag/SiO$_x$/VAMoS$_2$/Au devices.**

(a,b) 30 *I-V* switching cycles of a 60 nm and a 10 nm SiO$_x$/VAMoS$_2$ device, respectively. The devices exhibit TS with a positive voltage applied to the top Ag electrode. $I_{CC}$ was set to 1 µA, and arrows 1 and 2 denote the voltage sweep direction. (c) Cross-sectional TEM image of the device after RS at low HRS after several RS cycles. The dotted blue rectangle is further enlarged on the right, showing Ag traces in the material stack. (d,e) Statistical distribution of switching voltages ($V_{t,on}$ and $V_{hold}$) for 60 nm and 10 nm SiO$_x$/VAMoS$_2$ devices, respectively. 30 RS cycles were evaluated per sample, with data extracted from (a) and (b). (f) Device-to-device distribution of $V_{t,on}$ and $V_{hold}$ for 10 devices with 60 nm SiO$_x$/VAMoS$_2$ and 20 devices with 10 nm SiO$_x$/VAMoS$_2$. (g) Cumulative probability distribution of $V_{t,on}$ and $V_{hold}$ in a 10 nm SiO$_x$/VAMoS$_2$ device over 384 RS cycles.

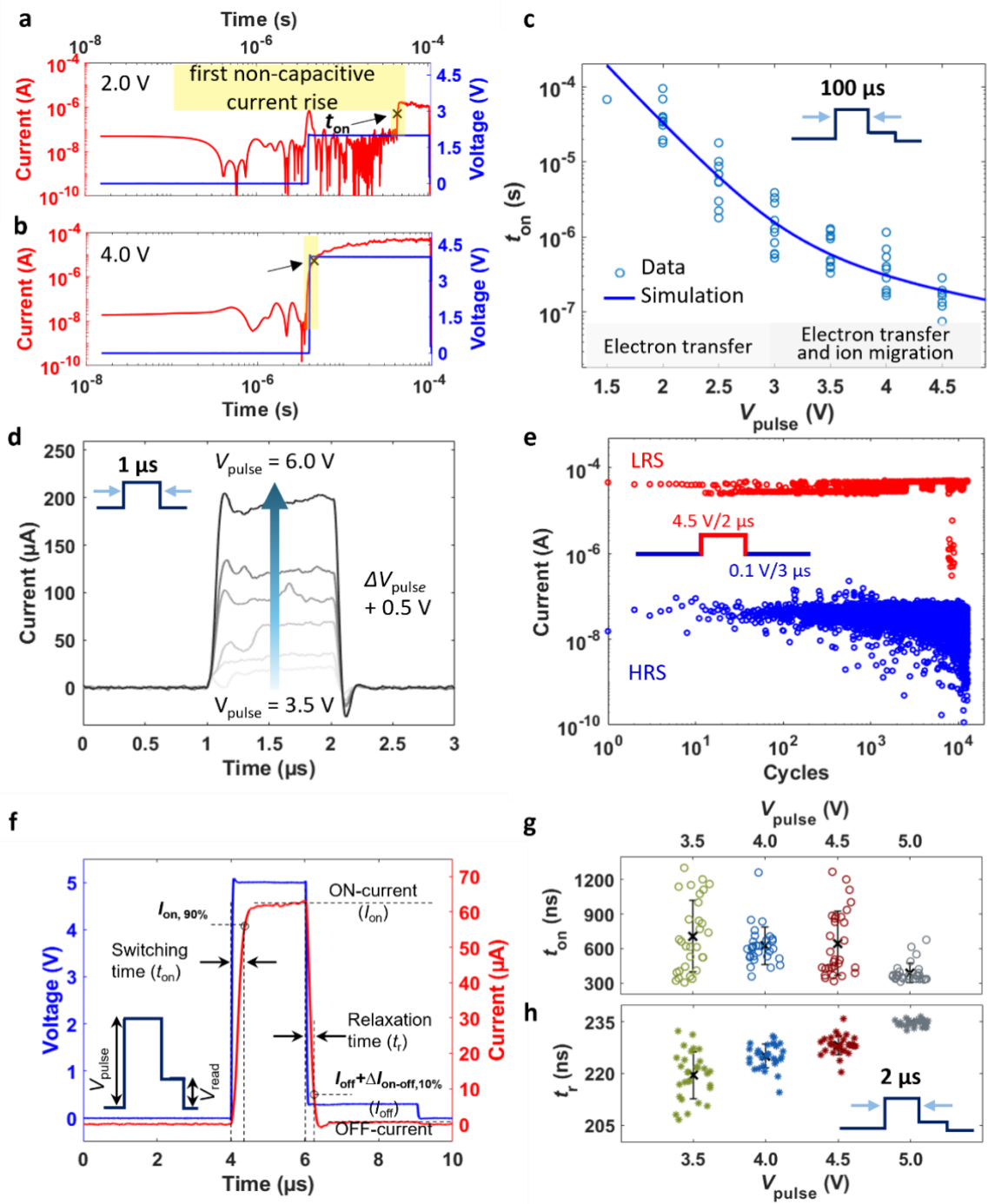

**Figure 4. Pulsed voltage characteristics of Ag/10 nm SiO$_x$/VAMoS$_2$/Au devices.**

(a,b) Transient current response under $V_{pulse}$ of 2.0 V (a) and 4.0 V (b) with a 100 μs duration in double logarithmic scale. $t_{on}$ is marked with an "x", and the first non-capacitive current rise is highlighted in yellow. (c) Switching kinetics showing experimental data for $t_{on}$ as a function of $V_{pulse}$ (blue circles) and fitting data from the numerical simulation (blue solid line). The plot is divided into two regions: electron-transfer-limited and mixed control regimes. The inset displays the voltage pulse waveform. (d) Transient current response of 1 μs voltage pulses with amplitudes ranging from 3.5 V to 6.0 V ($\Delta V_{pulse}$: +0.5 V). (e) Pulsed endurance showing over $10^4$ cycles. Inset: programming pulse consisting of 4.5 V/2 μs followed by 0.1 V/3 μs. The LRS and HRS currents were extracted as the average of the values recorded during the second half of the 4.5 V and 0.1 V pulses, respectively. (f) Current response in the time domain under a voltage pulse of 5.0 V/2 μs, followed by a read pulse 0.3 V/3 μs. The inset shows the corresponding voltage pulse waveform. (g,h) Statistical distributions of $t_{on}$ and $t_r$ as a function of $V_{pulse}$ in a short pulse of 2 μs for 30 repetitions, respectively.

**Table 1. Performance comparison of Ag-based threshold switches.**

| Device | Material growth method | $V_{t,on}$ | $V_{hold}$ | $I_{ON/OFF}$ (DC) | On-state current | | Endurance (cycles) | | Switching speed | | Pulsed voltage ($V_{pulse}$/duration) |
|---|---|---|---|---|---|---|---|---|---|---|---|
| | | | | | DC ($I_{cc}$) | AC | DC | AC | Switching time | Relaxation time | |
| **Ag/SiO$_x$/VAMoS$_2$/Au** This work | **CVD (Sulfurization)** | **0.63 V** | **0.5 V** | **> 10$^3$** | **1 µA** | **200 µA** | **384** | **>10$^4$** | **311 ns** | **233 ns** | **4.5 V/2 µs** |
| Ag/VAMoS$_2$/Au [12] | CVD (Sulfurization) | 0.35 V | 0.1 V | 10$^6$ | 100 µA | N/A | >30 | 5×10$^6$ | N/A | N/A | 1.2 V/50 µs |
| Ag/MoS$_2$/Ag [33] | Aerosol-Jet Printer (exfoliated flakes ink) | 0.18 V - 0.3 V | 0.13 V | 10$^5$ | 10 µA | 1 µA | 100 | N/A | N/A | 50 µs | 1.0 V/100 µs |
| Ag/DDG*/SiO$_2$/Pt [3] | magnetron sputtering | 0.6 V | - | 5x10$^8$ | 500 µA | 500 µA | 100 | 10$^6$ | 100 ns | 1 µs | 2.0 V/2 µs |
| Ag/SiO$_2$/Au [47] | N/A | 0.7 V | 0.05 V | N/A | 1 mA | 40 mA | N/A | N/A | N/A | 100 µs | 1.3 V/1 ms |
| Ag/SiO$_x$/C** [4] | Electron-beam evaporation | 1.75 V | 1.5 V | 10$^7$ | 50 µA | 15 µA | N/A | N/A | 10 µs | 1 ms | 5.0 V/100 µs |

Note: N/A, not available. DDG*: discrete-defect graphene. C**: graphitic carbon.

# Supporting Information

**Threshold Switching in Vertically Aligned MoS$_2$/SiO$_x$ Heterostructures based on Silver Ion Migration**

*Jimin Lee [1], Rana Walied Ahmad [2], Sofía Cruces [1], Dennis Braun [1], Lukas Völkel [1], Ke Ran [3,4,5], Joachim Mayer [4,5], Stephan Menzel [2], Alwin Daus [1,6,*], and Max C. Lemme [1,3,*]*

[1] Chair of Electronic Devices, RWTH Aachen University, Otto-Blumenthal-Str. 25, 52074 Aachen, Germany.

[2] JARA-Fit and Peter Grünberg Institute (PGI-7), Forschungszentrum Jülich GmbH, Wilhelm-Johnen-Str., 52428 Jülich, Germany.

[3] AMO GmbH, Advanced Microelectronic Center Aachen, Otto-Blumenthal-Str. 25, 52074 Aachen, Germany.

[4] Central Facility for Electron Microscopy, RWTH Aachen University, Ahornstr. 55, 52074 Aachen, Germany.

[5] Ernst Ruska-Centre for Microscopy and Spectroscopy with Electrons, Forschungszentrum Jülich GmbH, Wilhelm-Johnen-Str., 52428 Jülich, Germany.

[6] Institute of Semiconductor Engineering, University of Stuttgart, Pfaffenwaldring 47, 70569 Stuttgart, Germany.

\* Corresponding authors: max.lemme@eld.rwth-aachen.de and alwin.daus@iht.uni-stuttgart.de

**Section 1 Microscope Images of SiO$_x$/VAMoS$_2$ Cross-Point Devices.**

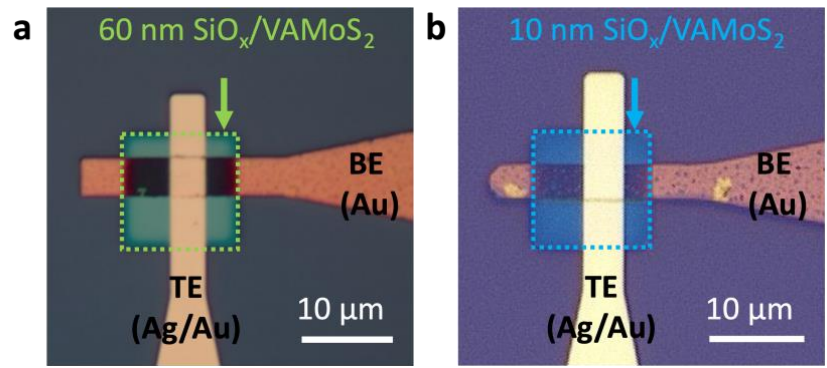

**Figure S1. Ag/SiO$_x$/VAMoS$_2$/Au devices.** (a,b) Top-view optical microscope images of SiO$_x$/VAMoS$_2$ cross-point devices with SiO$_x$ layer thicknesses of 60 nm (a) and 10 nm (b), respectively.

**Section 2. Raman Spectroscopy Characterization.**

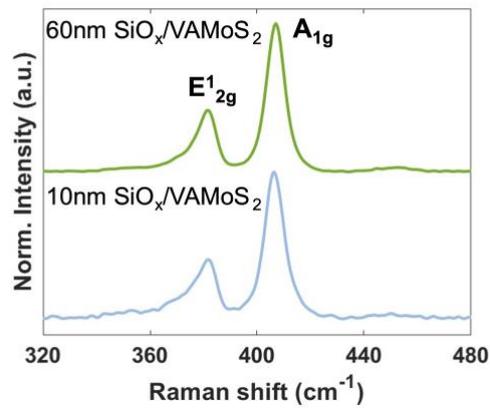

**Figure S2. Characterization of vertical orientation of MoS$_2$**. Raman spectra showing the E$^1_{2g}$ and A$_{1g}$ peaks at 382 cm$^{-1}$ and 407 cm$^{-1}$, confirming the 2H-MoS$_2$ phase in the 60 nm SiO$_x$/VAMoS$_2$ device (top) and the 10 nm SiO$_x$/VAMoS$_2$ device (bottom) on the Au surface, respectively.

**Section 3. Material Characterization Using HAADF and EDX.**

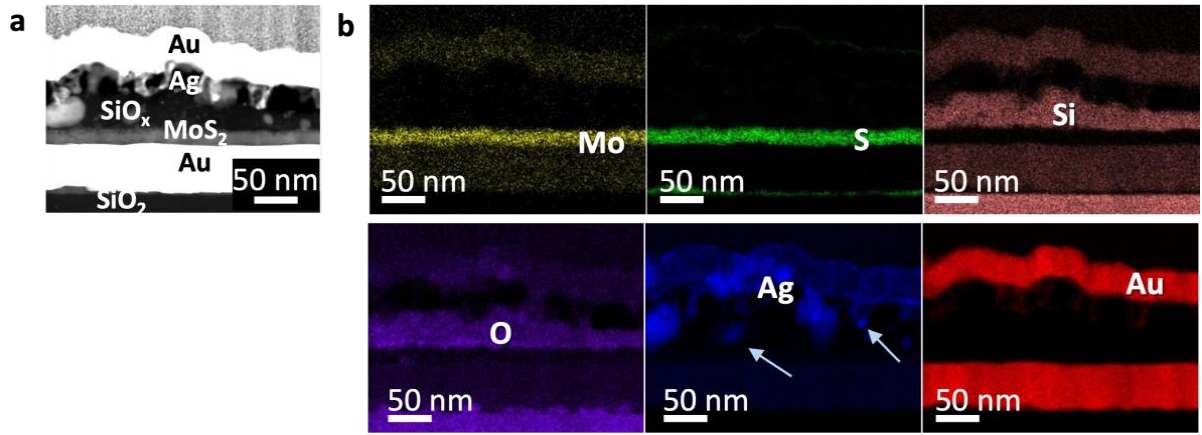

**Figure S3. Device structure of Ag/SiO$_x$/VAMoS$_2$/Au device with 60 nm SiO$_x$ layer.** (a) Cross-sectional HAADF image of the 60 nm thick SiO$_x$/VAMoS$_2$ device. (b) The corresponding EDXS elemental mapping of the device stack for Mo, S, Si, O, Ag, and Au. Compared to the 10 nm SiO$_x$/VAMoS$_2$ device, the SiO$_x$ layer thickness is less uniform and much thicker, which is attributed to the higher variability in volatile RS behavior. The volatile RS originated from the formation of Ag-diffused conductive filaments in the SiO$_x$/VAMoS$_2$ layer, as traces of Ag ions were observed in the SiO$_x$ layer.

## Section 4. Device to Device Variability.

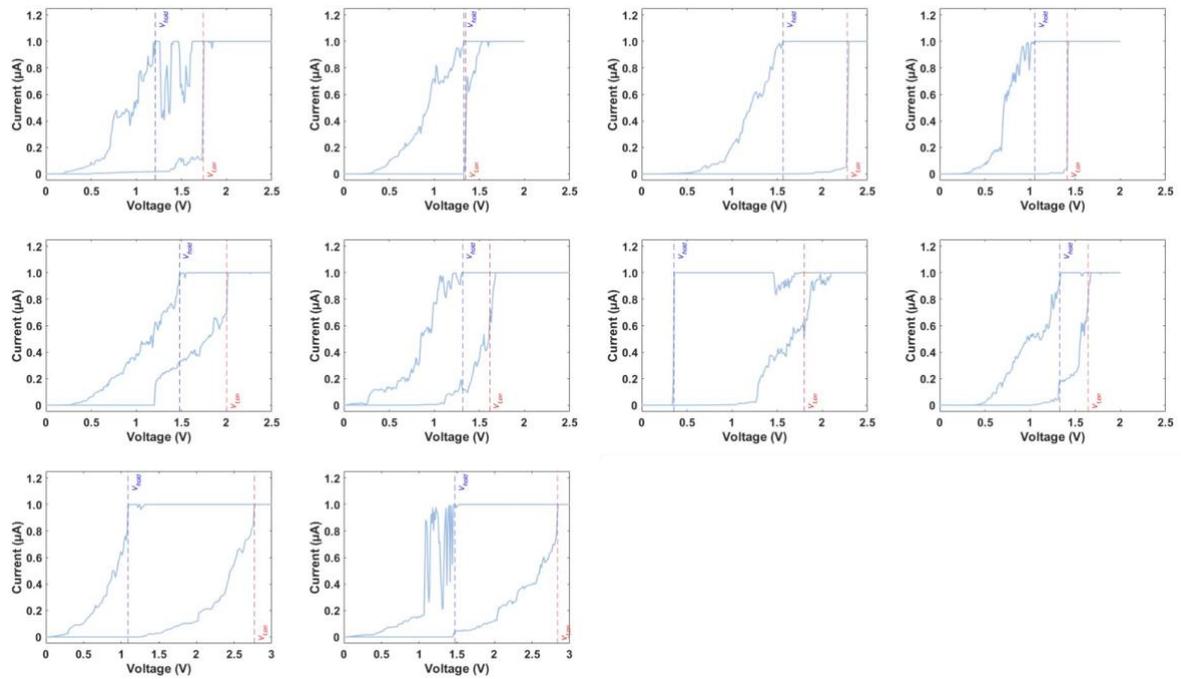

**Figure S4. Initial *I–V* sweeps of ten randomly selected 60 nm SiO$_x$/VAMoS$_2$ devices.** Raw data for $V_{t,on}$ (red dotted line) and $V_{hold}$ (blue dotted line) used in Figure 3f.

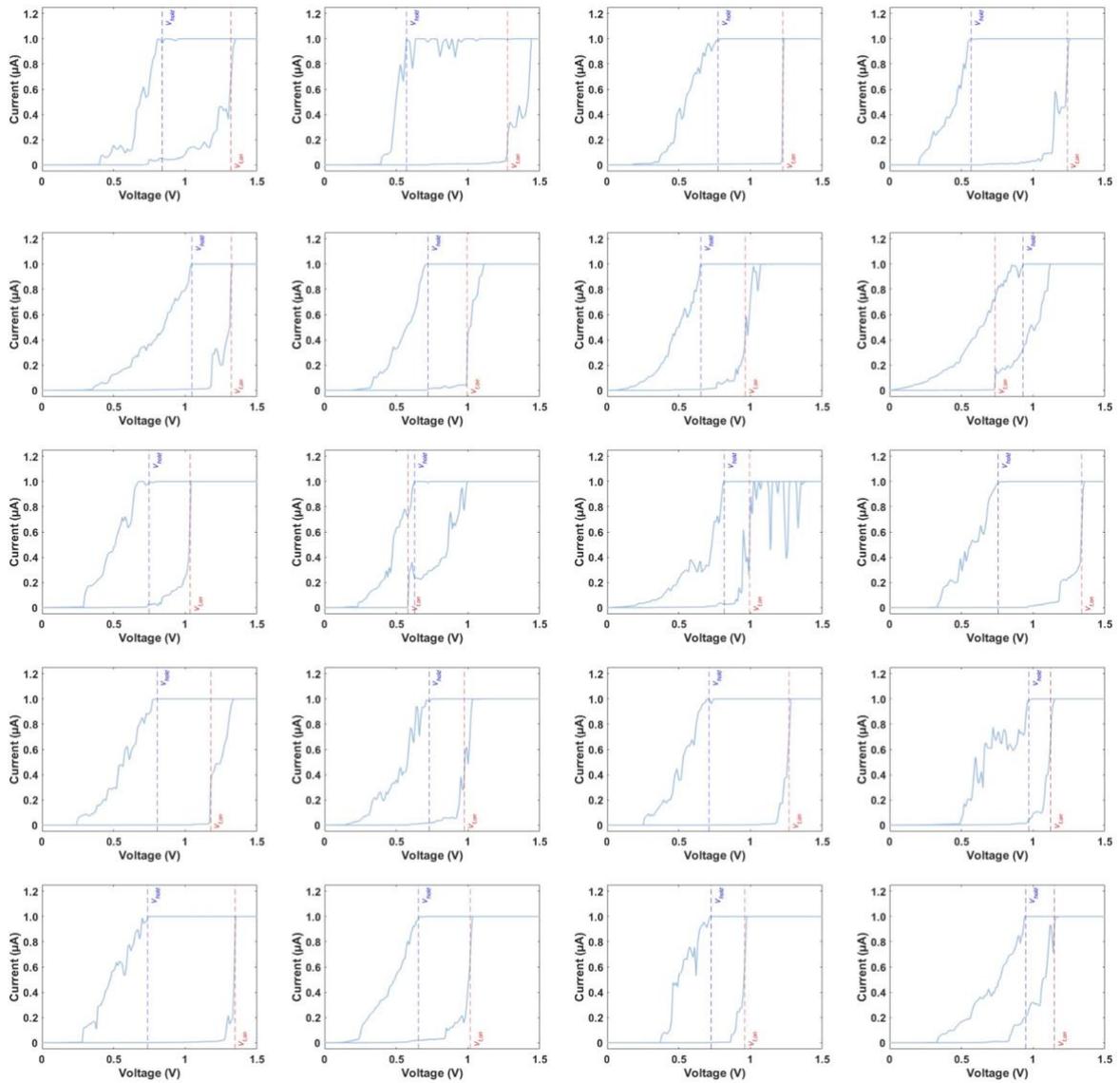

**Figure S5. Initial *I–V* sweeps of twenty randomly selected 10 nm SiO$_x$/VAMoS$_2$ devices.** Raw data for $V_{t,on}$ (red dotted line) and $V_{hold}$ (blue dotted line) used in Figure 3f.

**Section 5. Cyle to Cycle Variability.**

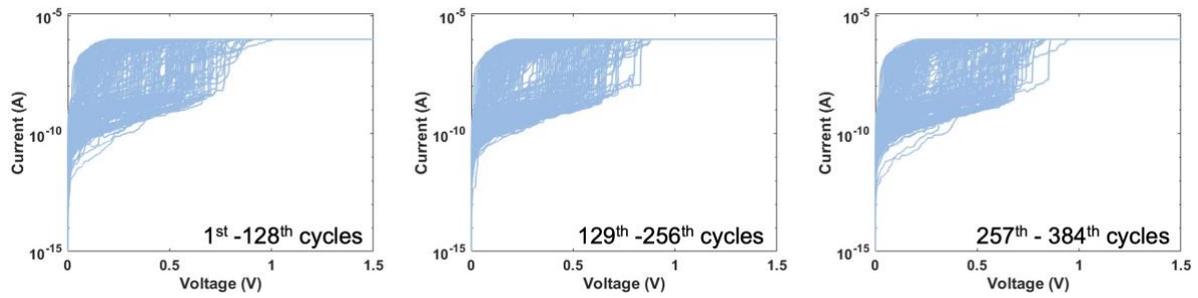

**Figure S6.** Raw data of 384 subsequent *I-V* sweeps from a 10 nm $SiO_x$/VAMoS$_2$ device.

**Section 6. Typical Dynamic Response During Pulsed Voltage Measurement With a 100 μs Pulse.**

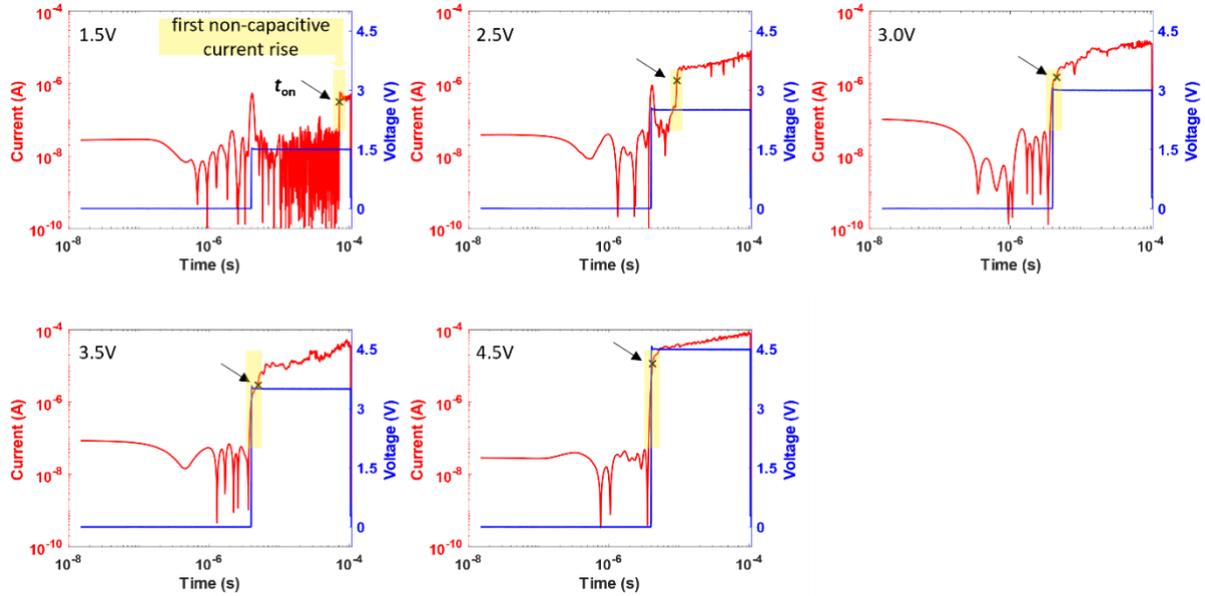

**Figure S7. Transient current response during the pulsed voltage measurement under different $V_{pulse}$ values ranging from 1.5 V to 4.5 V with a 100 μs duration in double-logarithmic scale.** The on-switching time ($t_{on}$) is marked with an "x", and the first non-capacitive current rise is highlighted in yellow. The transient current responses corresponding to $V_{pulse}$ values of 2.0 V and 4.0 V are shown in Figures 4a and 4b, respectively.

**Section 7. A Physics-based Dynamical Model for Switching Kinetics in the Conductive Filament Mechanism.**

The JART ECM model[1,2] describes the switching of the device into the LRS based on the filament length evolution upon applying an electrical stimulus. This filament evolution starts from the inert electrode and grows towards the active electrode throughout the switching layer. The state variable is the tunneling gap $x$ between the filament tip and the active silver electrode. The transient change of the gap $x$ is based on the ordinary differential equation

$$\frac{\partial x}{\partial t} = -\frac{M_{Me}}{ze\rho_{m,Me}} j_{ion}, \qquad (1)$$

where $M_{Me}$ is the molecular mass of silver, and $\rho_{m,Me}$ is its mass density, $z$ is the charge number, $e$ is the electron charge, and $j_{ion}$ represents the ionic current density contributions.

Upon applying a positive voltage to the active Ag electrode, electron transfer processes take place, here the oxidation of Ag atoms, causing the ionic Butler-Volmer current $I_{ac}$ at the interface of the active electrode and the switching layer. Besides the ionic migration/hopping process of these ions towards the inert electrode causing the ion hopping current $I_{hop}$, electron transfer processes as ionic Butler-Volmer current also take place at the switching layer/filament tip interface ($I_{fil}$), namely reduction processes that deionize the arriving Ag ions and contribute to a filament growth.

The electron transfer processes at the respective interfaces are described by a simplification of the Butler-Volmer equation. The latter equation in its original form, which describes these redox processes, is displayed by the following mathematical relation:

$$I_{ac/fil} = \pm zeck_{0,et}A_{ac/fil}\exp\left(-\frac{\Delta G_{et}}{k_BT}\right)\cdot\left[\exp\left(\frac{(1-\alpha_{et})ze}{k_BT}\eta_{ac/fil}\right) - \exp\left(-\frac{\alpha_{et}ze}{k_BT}\eta_{ac/fil}\right)\right] \qquad (2)$$

Here, the Ag ion concentration is given by *c*, $k_{0,et}$ is the electron transfer reaction rate, $A_{ac/fil}$ is the active electrode reaction area and the filament area, respectively, $\Delta G_{et}$ is the electron transfer activation barrier, $k_B$ is the Boltzmann constant, *T* is the temperature, $\alpha_{et}$ is the electron transfer coefficient, and $\eta_{ac/fil}$ is the respective electron transfer overpotential.

The ion migration current is given by the Mott-Gurney law:

$$I_{\text{hop}} = 2zecaf \exp\left(-\frac{\Delta G_{\text{hop}}}{k_B T}\right) A_{\text{is}} \sinh\left(\frac{aze}{2k_B T} \frac{\eta_{\text{hop}}}{x}\right) \quad (3)$$

Here, *a* describes the mean ion hopping distance, *f* is the hopping attempt frequency, $\Delta G_{hop}$ denotes the ion hopping activation barrier, $A_{is}$ is the cross-sectional area of the ion migration in the switching layer and $\eta_{hop}$ is the ion hopping overpotential. For the stack investigated in this work, these ion migration parameter values shall be considered as averaged values over the two inter-electrode layers, namely $SiO_x$ and $VAMoS_2$, as averaged values suffice for the main purpose of this analysis, which is to distinguish the limiting processes. In a future study, an individual description and modeling of each of the two layers and their interface might describe the dependence on different layer lengths and different strengths of migration limitation in a more precise way.

The nucleation process is an extended version of the electron transfer reduction process at the inert electrode, it occurs when the very first silver ions arrive at the inert electrode, and it is necessary to build a stable nucleus consisting of some Ag atoms that pave the way for the onset of Ag filament growth on top of the electrode. This and the aforementioned electron transfer and ion migration processes act as driving forces for the switching event and at least one of them limits the speed of this event as the rate-limiting process upon applying a certain voltage. In this physics-based dynamical model, one of the corresponding ionic currents from eq. (2) and eq. (3) is inserted into the description of the tunneling gap evolution from eq. (1) as the corresponding ionic current density to achieve the switching event. Once a sufficiently small gap *x* is achieved, the electronic tunneling current becomes the dominant current and mainly describes the LRS[1,2].

**Table S1. Model Parameters, Description and Values**[1,2]**.**

| Symbol/ Parameter | Value | Symbol/ Parameter | Value |
|---|---|---|---|
| $M_{me}$: molecular mass of silver | $1.7912 \cdot 10^{-25}$ kg | $r_{ac}$: radius of active electrode reaction area | 10 nm |
| $z$: charge transfer number | 1 | $r_{fil}$: filament radius | 1 nm |
| $\rho_{m,me}$: mass density of silver | $10.49 \cdot 10^3$ kg/m$^3$ | $r_{is}$: cross-sectional radius of ion hopping process | 1 nm |
| $m_r$: relative electron mass | 0.86 | $L$: length of switching/ insulating layer | 30 nm |
| $\Delta W_0$: eff. tunneling barrier height | 3.6 eV | $\rho_{fil}$: filament's electronic resistivity | $1.59 \cdot 10^{-8}$ Ωm |
| $\alpha_{et}$: electron transfer coefficient | 0.1 | $R_{el}$: resistance of electrodes | 40 Ω |
| $k_{0,et}$: electron transfer reaction rate | $3 \cdot 10^5$ m/s | $R_S$: series resistance | 120 Ω |
| $c$: Silver ion concentration | $2 \cdot 10^{26}$ 1/m$^3$ | $T$: temperature | 298 K |
| $\Delta G_{et}$: electron transfer activation barrier | 0.55 eV | $t_{0,nuc}$: prefactor of nucleation time | $2 \cdot 10^{-10}$ s |
|  |  | $\Delta G_{nuc}$: nucleation activation energy | 0.8 eV |
| $a$: mean ion hopping distance | 0.4 nm | $N_c$: number of atoms to achieve stable nucleus | 3 |
| $f$: ion hopping attempt frequency | $1 \cdot 10^{13}$ Hz | $\alpha_{nuc}$: electron transfer coefficient for nucleation | 0.5 |
| $\Delta G_{hop}$: ion migration barrier | 0.12 eV |  |  |

**Section 8. Pulsed Endurance of the Ag/SiO$_x$/VAMoS$_2$/Au Device With 10 nm SiO$_x$.**

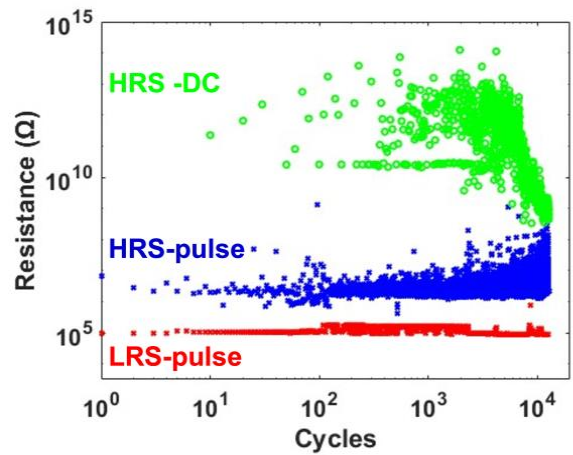

**Figure S8. Extracted LRS and HRS resistance values from pulsed measurements, with DC measured HRS resistance values.** The LRS and HRS resistance values from the pulse measurements (LRS-pulse and HRS-pulse) were determined by dividing the average applied voltage by the average on-state and off-state currents ($I_{on}$ and $I_{off}$), respectively. The values were taken during the second half of the pulse width for the 4.5 V and 0.1 V pulses. The HRS resistance values from the DC measurements (HRS-DC) were obtained by applying a 0.1 V pulse after every 10 cycles over a total of 10$^4$ cycles. Note that the HRS-pulse values were lower than the HRS-DC values. This discrepancy is attributed to the limited resolution of the pulse measurement setup (Keithley 4200S semiconductor parameter analyzer)[3], which depends on the current range, resulting in limited ability to determine HRS values.

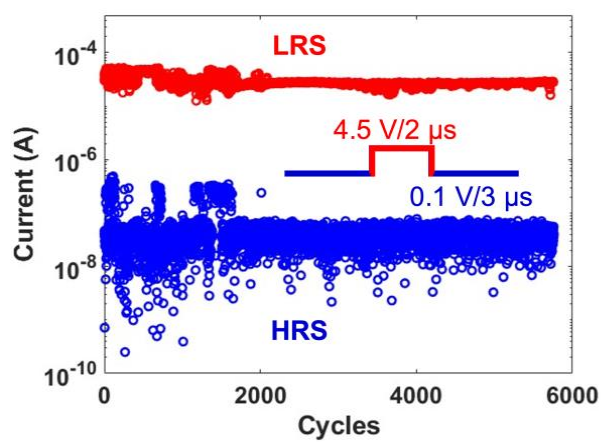

**Figure S9.** Pulsed endurance test showing ~5800 cycles without DC measurement. Inset: programming pulse consisting of 4.5 V/2 μs followed by 0.1 V/3 μs.

**Section 9. Comparison of Pulsed Voltage Characteristics of the Ag/SiO$_x$/VAMoS$_2$/Au Device and Ag/SiO$_x$/Au Device.**

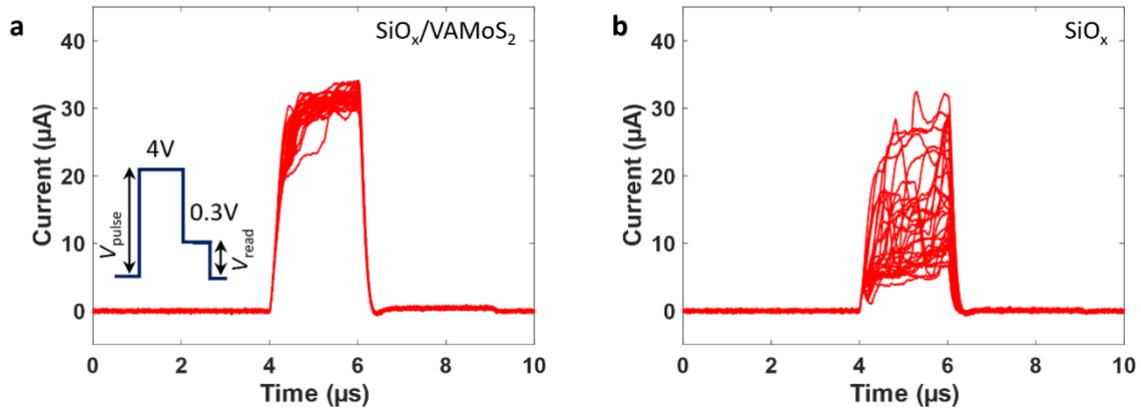

Figure S10. Repeatable transient current responses. Measured current waveforms over 30 consecutive cycles under identical applied voltage pulses for the SiOx/VAMoS2 device (a) and the SiOx-only device (b), respectively. Inset: programming pulse consisting of 4.0 V/2 μs followed by 0.3 V/3 μs.